\documentclass[iop]{emulateapj}
\usepackage{natbib,amssymb,amsmath,graphicx}
\newcommand\tna{\,\tablenotemark{a}}
\newcommand\tnb{\,\tablenotemark{b}}
\newcommand\tnc{\,\tablenotemark{c}}
\newcommand\tnd{\,\tablenotemark{d}}
\newcommand{\HH}{$\mathrm{H}_{2}$ }

\newcommand{\Brg}{Br$\gamma$ }
\newcommand{\UY}{UY~Aur }
\newcommand{\UYA}{UY~Aur~A }
\newcommand{\UYB}{UY~Aur~B }

\begin{document} 

\title{Variable accretion processes in the young binary-star system UY
Aur$^{*}$}\thanks{$^{*}$Observations reported here were obtained at the MMT
Observatory, a joint facility of the University of Arizona and the Smithsonian
Institution.}

\author{Jordan M. Stone, J.~A. Eisner}

\affil{Steward Observatory, 
University of Arizona,
933 N. Cherry Ave,
Tucson, AZ 85721-0065, USA;
jstone@as.arizona.edu, jeisner@as.arizona.edu}

\author{Colette Salyk}

\affil{National Optical Astronomy Observatory, 
950 North Cherry Avenue, Tucson, 
AZ 85719, USA; csalyk@noao.edu}

\author{Craig Kulesa, Don McCarthy}

\affil{Steward Observatory, 
University of Arizona,
933 N. Cherry Ave,
Tucson, AZ 85721-0065, USA;
dmccarthy@as.arizona.edu, ckulesa@as.arizona.edu}

\begin{abstract} 

We present new K-band spectroscopy of the UY Aur binary star system. Our data
are the first to show \HH emission in the spectrum of UY~Aur~A and the first to
spectrally resolve the \Brg line in the spectrum of UY~Aur~B.  We see an
increase in the strength of the \Brg line in UY~Aur~A and a decrease in \Brg
and \HH line luminosity for UY~Aur~B compared to previous studies.  Converting
\Brg line luminosity to accretion rate, we infer that the accretion rate onto
\UYA has increased by $2\times10^{-9}$ M$_{\odot}$yr$^{-1}$ per year since
a rate of zero was observed in 1994.  The \Brg line strength for \UYB has
decreased by a factor of 0.54 since 1994, but the K-band flux has increased by
0.9 mags since 1998. The veiling of UY~Aur~B has also increased significantly.
These data evince a much more luminous disk around UY~Aur~B. If the lower \Brg
luminosity observed in the spectrum of \UYB indicates an intrinsically smaller
accretion rate onto the star, then \UYA now accretes at a higher rate than
UY~Aur~B. However, extinction at small radii or mass pile-up in the
circumstellar disk could explain decreased \Brg emission around \UYB even when
the disk luminosity implies an increased accretion rate.  In addition to our
scientific results for the UY~Aur system, we discuss a dedicated pipeline we
have developed for the reduction of echelle-mode data from the ARIES
spectrograph.

\end{abstract}

\section{Introduction} 
A large fraction ($\sim 30\%$) of of all main sequence stars in the galactic
disk are in binary or multiple systems \citep[e.g.][]{Lada06}. The
main-sequence binary fraction is a function of spectral type, varying from near
100\% for early type stars down to $\sim20\%$ for late M-type stars
\citep[][and references therein]{Raghavan10}. While observations of rich
star-forming clusters seem to exhibit a binary fraction consistent with the
main sequence value, loose star-forming associations show considerable excess
\citep[][and references therein]{Duchene99b}. In fact, surveys of the young
low-mass stars in the Taurus-Auriga star-forming region indicate a binary
fraction of $\gtrsim 75\%$ \citep{Ghez93,Leinert93,Kraus11}. Furthermore, young
classical T-Tauri star binaries appear to be more co-eval than random pairings
of stars within a cloud, implying the binary formation mechanism operates
within $\sim1$ Myr \citep{Prato97, Duchene99b, White01, Hartigan03}. The high
incidence of binaries in loose star forming regions implies that formation in
multiple systems is the norm \citep[see e.g.][for a recent review]{Duchene13}.
While the multiplicity statistics of dense star-forming regions do not
show significant excess over the main-sequence, it is still undetermined
whether stars form with different multiplicity statistics in these
environments, or whether shorter dynamical times accelerate the disruption of
multiple systems \citep{Duchene13}.  

It is important to understand the details of the binary-star formation process
given the significance of binary-star formation both as a substantial
contributor to the galactic stellar population and in influencing planet
formation and stellar feedback. 

Theoretical studies have elucidated how dynamical resonances and tidal torques
conspire to open a gap in a viscous circumbinary disk \citep[e.g.][]{Lin79}.
Matter can accrete through such gaps in geometrically thin streams
\citep[e.g.][]{Artymowicz96}. Studies of the flow from the circumbinary disk
onto the circumstellar disks indicate that the primary star should have the
more massive disk, however, model predictions of which star is preferentially
fed new material seem to depend on the adopted viscosity
\citep[e.g.][]{Bate97,Ochi05}. Recently sub-mm observations of young binary
stars have revealed some systems where the secondary is host to the most
massive disk, possibly implicating multiple formation modes for binary stars
\citep{Akeson14}.

UY~Aur is a Classical T-Tauri star binary system. The stars are separated by 0.9$''$
($\sim120$ AU).  When \citet{Joy44} first identified UY~Aur as a binary in the
optical, the authors reported a $\Delta\mathrm{m}\approx0.5$.  A few decades
later, \citet{Herbst95} could not detect the secondary and placed a lower limit
of $\Delta\mathrm{m}\approx5$ at R-band. UY~Aur has since been observed as
a binary in the infrared \citep[e.g.][]{Ghez93, Leinert93, Close98,
Brandeker03, Hioki07}.  Resolved near-infrared photometry by \citet{Close98},
\citet{Brandeker03}, and \citet{Hioki07} demonstrated that the flux ratio is
variable in the infrared as well.  Combining the H-band measurements of
\citet{Close98}, and \citet{Hioki07}, UY~Aur A varied relatively little,
changing by $\lesssim 0.4$ magnitudes over 3 epochs in 10 years. UY~Aur B, on
the other hand, varied by up to 1.3 magnitudes in the same time. 

\citet{Herbst95} performed spatially resolved K-band spectroscopy of both
sources. Those authors showed that \UYB exhibited strong \Brg and \HH v=1-0
S(1) emission lines, while \UYA showed none. \Brg is a canonical accretion
tracer, and their observations showed that UY~Aur A can spend some time in
a state with a very low accretion rate. The presence of \HH v=1-0 S(1) emission
without \HH v=2-1 S(1) emission led \citet{Herbst95} to believe it was arising
in a shock. They suggested that the shock could arise from a stellar wind
originating from \UYA and impacting circumstellar material around \UYB. 

Millimeter molecular line observations revealed a Keplerian disk surrounding
the stars \citep{Dutrey96, Duvert98}. The disk has an inner rim of $\sim500$
AU \citep{Hioki07}. By studying the reflected infrared light from the circumbinary disk,
\citet{Close98} was able to deduce a small/unrefined dust grain distribution.
When \citet{Skemer2010}---using spatially resolved N-band spectroscopy--- found
a similar distribution of dust grains in the circum\textit{stellar} disk of
UY~Aur~A, they noted that either something is prohibiting the evolution of
grains in the stellar disk, or the grains are being replenished from the
circumbinary reservoir, a clue that material is making its way from the
circumbinary disk to the circumstellar disks, as predicted by theory.

Throughout this paper we assume the stellar spectral types and masses deduced
by \citet{Hartigan03}. Using STIS on board $HST$, \citet{Hartigan03}
obtained spatially resolved medium and low resolution spectra from
$0.5$--$1\mu$m. They simultaneously fit their data for veiling, spectral type,
and reddening. Their best fit for \UYA is a 0.6 M$_{\odot}$ M0 type star with
A$_{v}$=0.55. For UY~Aur~B, their best fit parameters indicate a 0.3 M$_{\odot}$,
M2.5 type star with A$_{v}$=2.65. The significant difference in extinction
toward each source is uncommon among their sample, but is consistent with the
large change in observed flux of \UYB between the 1940's and 1990's. In fact,
variable reddening has been suggested as the driver of the NIR variability in
\UYB \citep[e.g.][]{Brandeker03}.

Here we report new spatially and spectrally resolved observations of UY~Aur. We
present the first detection of \HH lines in the spectrum of UY~Aur~A, and the
first spectrally resolved measurement of \Brg in the spectrum of UY~Aur~B. We
highlight a significant change in the relative strengths of both the \Brg line
and the \HH v=1-0 S(1) line between our observations and the only other
published K-band spectrum of UY~Aur~B \citep{Herbst95}.  

\section{Observations and Reduction} \label{ObsSec} 

We observed both components of UY~Aur on 3 October 2012 UT using ARIES on the
MMT. ARIES is optimized for diffraction limited observations behind the MMT
adaptive optics system. We used the f/5.6 (0.1$''$/pixel) echelle mode with
a 1$''$x0.2$''$ slit.  This mode is designed to provide $\gtrsim20$ cross-dispersed
spectral orders at a resolution of R$\approx30,000$. Our setup targeted
a wavelength range of $\sim1.6\mu\mathrm{m}$ -- $2.47~\mu\mathrm{m}$ with some
gaps between orders.  The adaptive optics provided 0.2$''$ FWHM PSFs on the
imaging side of ARIES and we realized 0.4$''$ FWHM spatial profiles on the
spectrograph side. The degraded spatial profile is most likely due to one of
the optics on the spectrometer side of ARIES. Measured argon-lamp emission
lines indicate a realized spectral resolution of R=15,000.

Each component of the UY~Aur binary was observed with the slit aligned along
two anti-parallel position angles: $40^{\circ}$ and $-140^{\circ}$. This was
done in order to help characterize instrumental effects on the spectral traces
\citep[e.g.][]{Brannigan06}. An observing log is included in Table
\ref{obsTable}. We slewed off source to collect sky frames once for the pair of
PA=$40^{\circ}$ observations and once for the pair of PA=$-140^{\circ}$
observations. Likewise, observations of the A7 star HR 1620 were made once for
each position angle to facilitate the correction of telluric absorption
features in our data. Flat field frames were collected to correct for
inter-pixel variations in the detector gain.  

To monitor instrumental flexure and spectral fringing effects, we collected
flat field frames at each new pointing during the night. Dark frames to correct
the flats were collected at the end of the night. 

\begin{figure}[h!]
\begin{center}
\epsscale{0.75}
\plotone{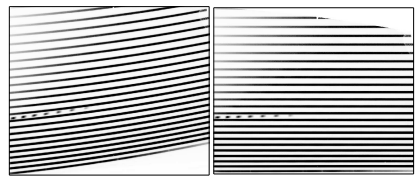}
\caption{Dewarping a flat-field frame. On the left, we show an original warped
image of a flat lamp with many curved orders. On the right we show the straight
spectral orders after dewarping.\label{flatFig}}
\end{center}
\end{figure}

\begin{figure}[h!]
\begin{center}
\epsscale{0.8}
\plotone{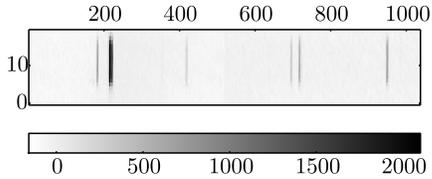}
\caption{ Argon arclamp emission lines are rectified after using a fifth-degree
two-dimensional polynomial transform.\label{argonFig}}
\end{center}
\end{figure}

We developed a dedicated pipeline for the reduction of echelle-mode data from
ARIES to analyze our observations. We began with subtracting sky frames from
science frames, and dark frames from flat frames.  We then dewarped the
spectral orders using a fifth-degree two-dimensional polynomial coordinate
transformation to map the curved stellar spectral traces to straight lines.
Figure \ref{flatFig} shows an example of dewarping a flat-field frame using the
transformation derived from the stellar traces.  We used a fifth-degree
transformation as this was adequate to ensure that the spatial dimension of
each order was perpendicular to the spectral dimension, as we verified using
observations of an Argon arclamp (Figure \ref{argonFig}).

\begin{figure}[h!]
\begin{center}
\includegraphics[width=0.5\textwidth]{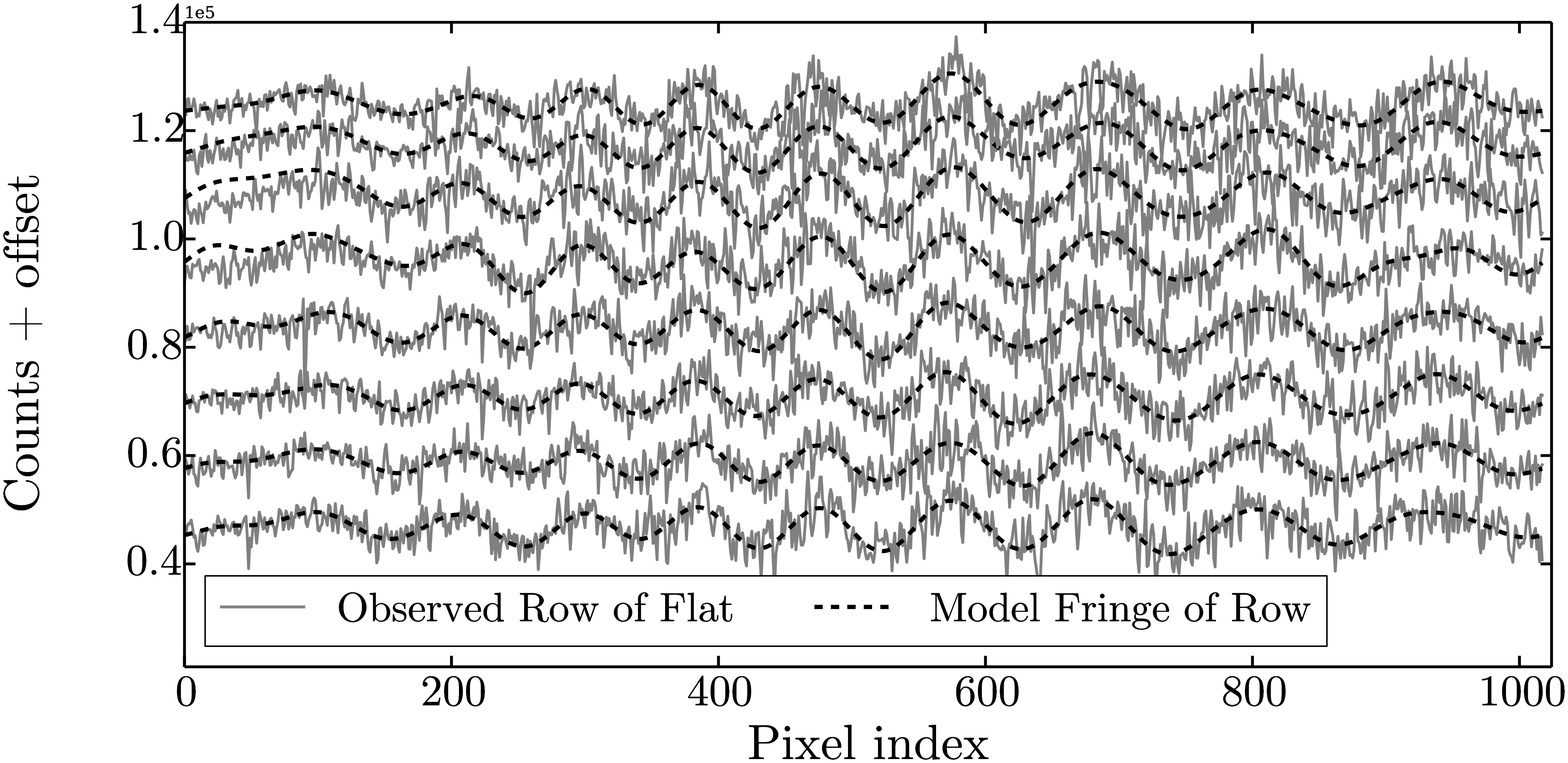}
\caption{This plot illustrates, for one order, our approach to defringing the flat-field
frames. A model fringe for each row was made using a low-pass
Fourier filter. These models were subtracted from the flat, and saved for use
in defringing the science targets. In this figure, the flux in each row is
plotted versus pixel number in gray. Each row is manually offset for clarity.
The model fringe is over-plotted with a dashed black line.\label{flatdefringe}}
\end{center}
\end{figure}

After dewarping all the frames, we began defringing, beginning with the flats.
The dominant frequency of the instrumental fringing is $\sim0.01~
\mathrm{pixels}^{-1}$. This corresponds to a fringe wavelength of $\sim 500$
km/s. Our flat-defringing algorithm proceeded row-by-row in each order as
follows. We first fitted and subtracted a third-order polynomial to the row. We
then created a model of the fringe pattern using a Fourier filter which only
passed frequencies less than 0.015 $\mathrm{pixels}^{-1}$ (see Figure
\ref{flatdefringe}).  Next, we subtracted this model fringe from the row it was
created from and then added back in the third-order polynomial that was
previously subtracted. This was done in order to remove the fringing but
maintain the very low frequency shape of the flat. We saved our derived model
fringe in a template for later re-use in defringing the stellar spectra.

After defringing the flats, we divided the flat frames into the stellar
spectra. Our next step was to extract from the two-dimensional spectral orders
one-dimensional spectra.  To extract the 1-d spectra we first fit a Gaussian to
the average spatial profile of the order. Then, for each column, we summed the
light in each row 1-$\sigma$ above and below the average spatial centroid. 

\begin{figure*}[H!]
\begin{center}
\epsscale{0.8}
\plotone{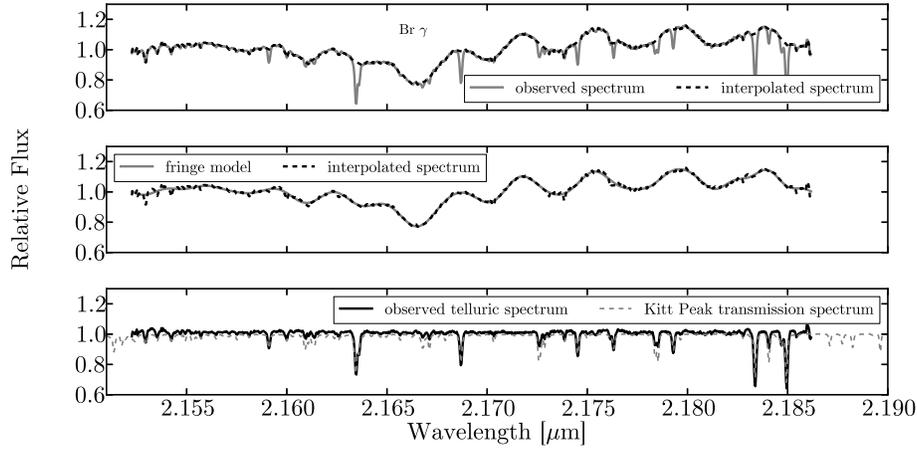}
\caption{In order to defringe the orders of the telluric calibrator spectrum we
first interpolated over narrow telluric absorption lines (top panel). We then
used a low-pass Fourier filter to generate a model fringe. Since our calibrator
is a fast rotator, this method also corrected for photospheric absorption
(middle panel). In the bottom panel we show the spectrum of HR 1620 after
subtracting the model fringe, which corrected for both the instrumental fringing
and the photospheric absorption. The technique of interpolating over sharp
features before generating a model fringe was also applied to the orders of the
UY~Aur spectra which included narrow \HH lines.\label{HR1620defringe}}
\end{center}
\end{figure*}

With the stellar spectra extracted, our next step was to defringe them.  We
started with the telluric calibrator. Since sharp features have broad Fourier
transforms, narrow telluric absorption lines in the telluric calibrator
spectrum can bias the instrumental fringe model when using even a low-pass
Fourier filter. To avoid this, before creating the model fringe, we first
interpolated over telluric lines. We have illustrated this step in the top
panel of Figure \ref{HR1620defringe}. We created a model fringe by first
subtracting a best-fit low-order polynomial and then using a Fourier filter to
pass only the spatial frequencies less than 0.015 $\mathrm{pixels}^{-1}$ (see
the middle panel of Figure \ref{HR1620defringe}).  Finally, we subtracted this
model fringe from the original spectrum to remove the instrumental fringing and
then added back in the best-fit polynomial.  Since our calibrator source, HR
1620, rotates quite quickly \citep[$v\mathrm{sin}i=131$ km/s,][]{Royer02}, our
Fourier filtering approach had the ancillary effect of removing rotationally
broadened photospheric lines. In the bottom panel of Figure
\ref{HR1620defringe}, we compare to the observed telluric transmission spectrum
provided by \citet{Hinkle02}. To facilitate the comparison, we show the
spectrum of HR 1620 before we replace the best-fit polynomial shape.

Next, we performed a wavelength calibration on both the defringed telluric
spectrum and the as-yet uncorrected target spectra. Wavelength calibration was
done by fitting a fourth-degree polynomial to transform measured telluric
absorption line pixel positions to the wavelengths of each line provided by
\citet{Hinkle02}. We then divided the telluric calibrator spectrum into the
target spectra to remove telluric absorption features. At this point, we summed
the spectra from each PA for each target, creating one higher signal-to-noise
spectrum for each source.

\begin{figure*}[h!]
\begin{center}
\epsscale{0.8}
\plotone{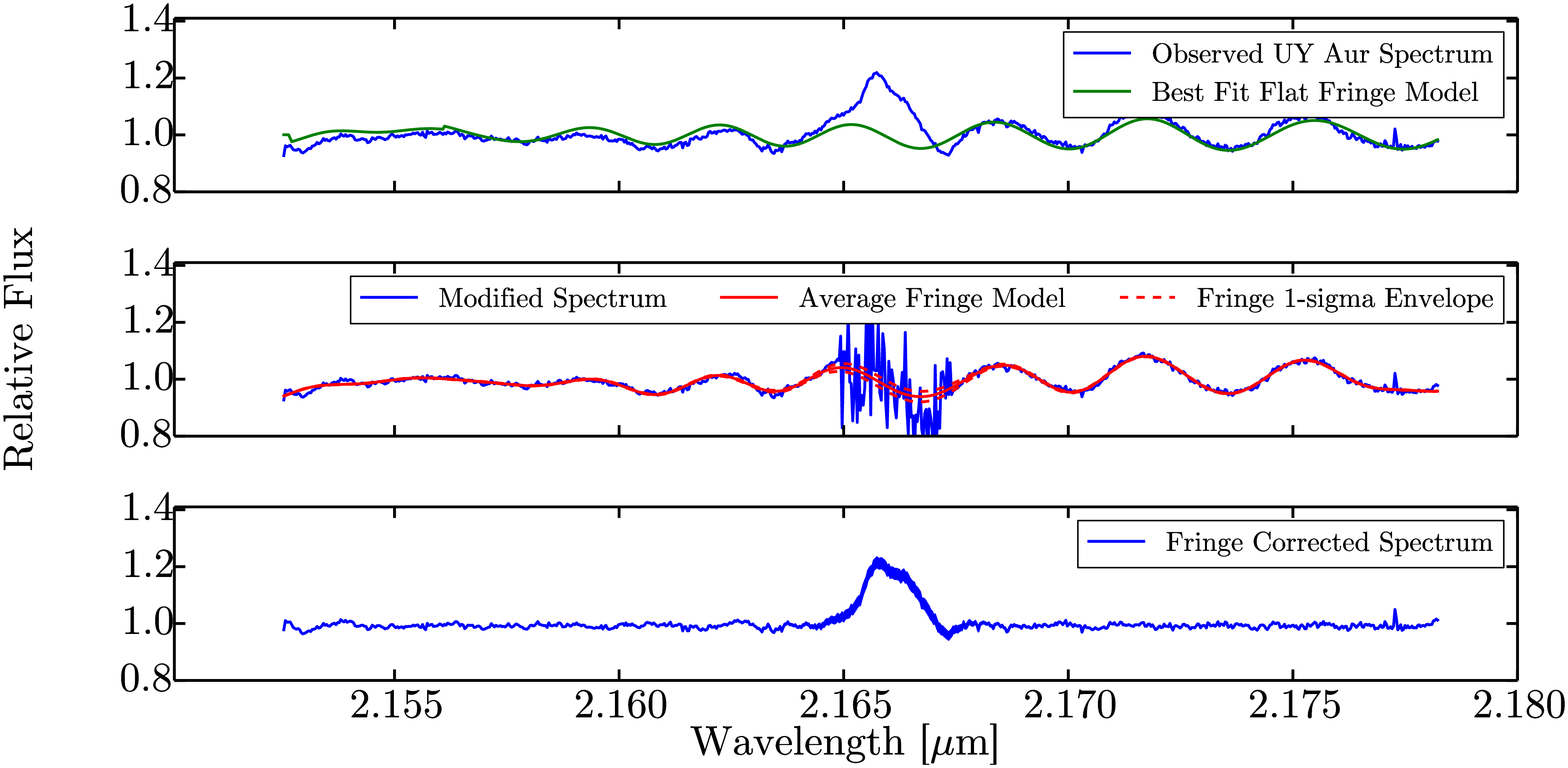}

\caption{Top: Fit of the 1-d flat-field fringe to the observed spectrum.
Middle: Regions of known broad-emission features are replaced with the best fit
flat fringe and noise. This is repeated 1000 times, generating a model fringe with
a Fourier filter each time. Bottom: The defringed spectrum, the width of the
swath indicating the range of allowed solutions given the precision of our
defringing model\label{UYAurdefringe}} 

\end{center} 
\end{figure*}

To defringe the target spectra, we interpolated over narrow emission lines and
photospheric absorption lines before defringing. For broad features, such as
the Br$\gamma$ emission line, we had to take care not to model out real
emission while still generating an accurate model for the instrumental
fringing. To do this, we first generated a best-guess model fringe. Our
best-guess fringe was created in two steps: 1) we extracted a one-dimensional
flat-field fringe by summing the same rows in the template of saved flat
fringes that were summed in the science frame during the extraction of the
science spectrum, and 2) we fit this one-dimensional flat-field fringe to the
target spectrum by adjusting the wavelength solution parameters to produce
a closest match. We show this step in the top panel of Figure
\ref{UYAurdefringe}. Next we replaced spectral regions of known features in our
observed spectra, with the corresponding region of our best-guess fringe
down weighted by the addition of Gaussian noise. We then used a low-pass Fourier
filter to generate a model fringe (middle panel of Figure \ref{UYAurdefringe}).
We repeated this process 1000 times, each time with a new realization of the
Gaussian noise, to produce 1000 model fringes. The average of the 1000 model
fringes we took as our fiducial fringe for defringing. The variation in the
1000 fringe models suggested the precision of our model fringe. In the bottom
panel of Figure \ref{UYAurdefringe}, we show a portion of the defringed spectrum of
UY~Aur~A in the vicinity of Br$\gamma$, plotted with a blue swath. The width
of the swath at each wavelength represents the 1-$\sigma$ variation in the
fringe model at that point.

\begin{deluxetable}{cccc} 
\tabletypesize{\scriptsize} 
\tablecolumns{4}
\tablewidth{0pt} 
\tablecaption{Observations} 
\tablehead{ \colhead{Source}& 
            \colhead{Total Exposure Time [s]} & 
            \colhead{Slit PA} & 
            \colhead{Airmass}
          } 
\startdata 
UY~Aur A   & 899.1 & $  40^{\circ}$ & 1.00  \\ 
UY~Aur B   & 899.1 & $  40^{\circ}$ & 1.01 \\ 
HR 1620    & 149.5 & $- 47^{\circ}$ & 1.03 \\ 
UY~Aur B   & 899.1 & $-140^{\circ}$ & 1.04 \\ 
UY~Aur A   & 799.2 & $-140^{\circ}$ & 1.07 \\ 
HR 1620    & 149.5 & $  56^{\circ}$ & 1.05 \\  
\enddata
\tablecomments{ Each member of the UY~Aur binary was observed twice. Each
observation was performed with anti-parallel slit orientations.  We use HR 1620
to perform the telluric calibration. After performing a wavelength calibration
and correcting for telluric lines in the spectra of UY Aur A and B, we sum over
rotation angle to produce a single spectrum for each source.  \label{obsTable}}
\end{deluxetable}

\section{Results and Analysis}\label{resultsSec} 
\subsection{Equivalent Widths}

\begin{figure}[h!]
\begin{center}
\epsscale{0.8}
\plotone{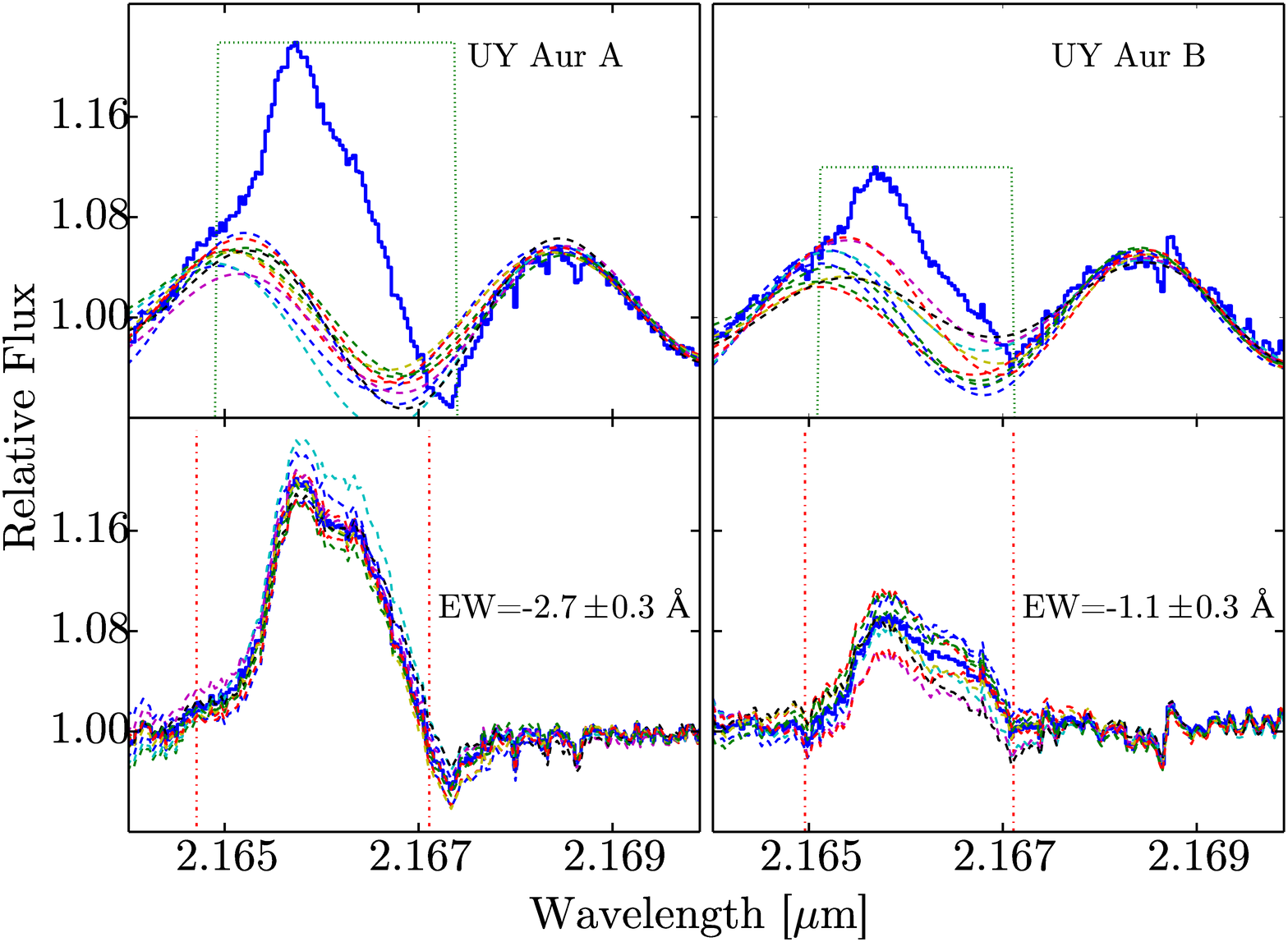}
\caption{Measuring the EW of \Brg. The left column corresponds to UY~Aur A, and
the right column corresponds to UY~Aur B. In the top panels, we show with
a blue histogram the observed spectrum in the vicinity of \Brg. The dotted
top hat indicates the portion of the observed spectrum which we excised and
replaced with a noisy first-guess fringe. The smooth dashed fringes show 10
examples of the 1000 model fringes derived using this method. In the bottom
panels we show the defringed spectrum for each of the 10
models to illustrate the range of our defringing confidence. This approach
introduces $\sim0.3$ \AA~error into our EW measurement.\label{EWfig}} 
\end{center} 
\end{figure}

We measured the equivalent width (EW) of the \Brg, \HH v=1-0 S(1), and \HH
v=1-0 S(0) lines for both sources. For the broad \Brg lines, we performed the
EW measurement during the repetitive fringe modeling process (see Section
\ref{ObsSec}). For each repetition, we generated and subtracted a new fringe
model from the observed spectrum and measured the EW. To illustrate this
process we show 10 repetitions in Figure \ref{EWfig}.  The mean and standard
deviation of all 1000 EW measurements is our adopted EW and uncertainty,
respectively.  The defringing process appears to introduce $\sim0.3$ \AA~of
uncertainty into our EW measurements for \Brg. We do not correct our measured
EWs for photospheric absorption, in line with previous studies
\citep{Herbst95,Fischer11}. Given the spectral types of the stars \citep[M0 and
M2 for UY Aur A and B, respectively][]{Hartigan03}, even with zero veiling the
effect on our measured EWs is 10\%, within our measurement uncertainty.

For the narrow \HH lines, we did not iterate to produce the fringe model
because the \HH lines are narrow compared to the fringing, and can be readily
distinguished (see Section \ref{ObsSec}). To estimate our precision in the
measurement of the EW of these lines, we ascribed to each pixel in the line the
same level of noise as measured in the nearby continuum. We then calculate the
EW uncertainty to be $\sqrt{N}\sigma$, where $N$ is the number of pixels in the
line and $\sigma$ is the measured noise level. All EW measurements are
presented in Table \ref{EWTable}.

\subsection{Line Luminosities} 

To convert our EW measurements to line luminosities we used D=140 pc
\citep{Elias78}, and the K-band photometry of \citet{Close98} for UY~Aur A.
For UY~Aur B, which has shown significant NIR variability
\citep[e.g.][]{Hioki07}, we determined the flux ratio to the primary using
imaging data taken at the same time as our spectra. Line luminosities are not
corrected for extinction. In Section \ref{Disc} we discuss how
variable extinction may effect the relative line luminosities. All resulting
emission line luminosities are reported in Table \ref{EWTable}.

\subsection{Accretion Rates}
We converted our measured \Brg line luminosity to accretion luminosity, without
correcting for extinction, using the correlation reported by
\citet{Muzerolle98}. Accretion rates were derived from accretion luminosities
using Equation 11 in \citet{Hartigan03} ---which assumes that accreting
material is in free fall from 3R$_{*}$ when it hits the stellar surface and
that half of the accretion energy is radiated directly, the other half heats
the star. Our accretion rate measurements are listed in Table \ref{EWTable}. 

\begin{figure}[h!]
\begin{center}
\epsscale{0.8}
\plotone{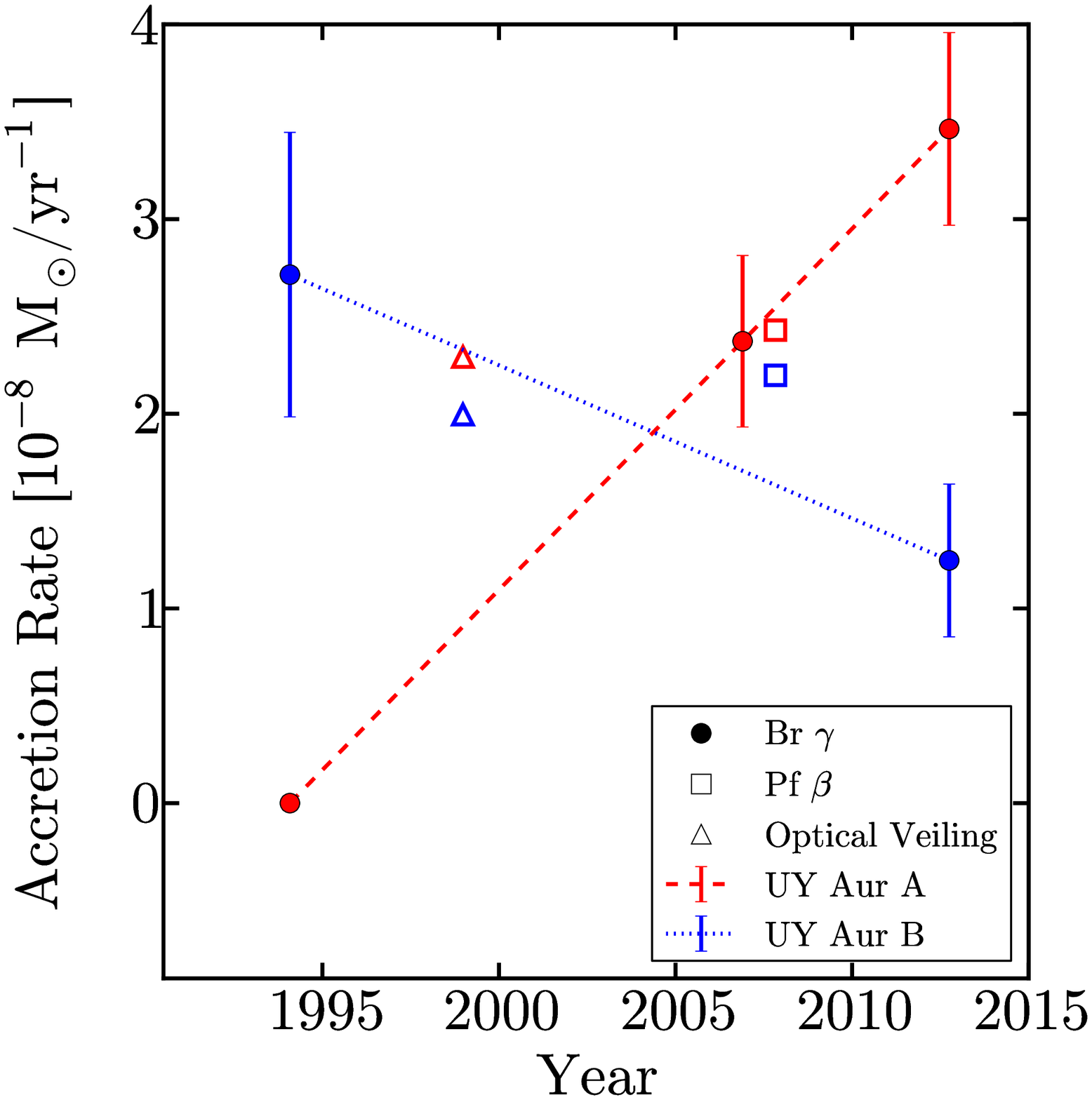}
\caption{The mass accretion rate versus time for each component of the binary.
Red symbols correspond to UY~Aur A, and blue symbols correspond to UY Aur B.
The filled circles, which have been connected by either a dashed (for UY~Aur~A)
or dotted (for UY~Aur~B) line, represent measurements made using \Brg emission
not corrected for extinction (\Brg data are from \citet{Herbst95},
\citet{Fischer11}, and this work).  Errorbars include contributions from
uncertainty in the continuum flux and in the EW.  The open triangles show the
accretion rate derived via a measurement of the optical veiling and are from
\citet{Hartigan03}. The open squares show the accretion rate measured using
Pf$\beta$ emission and are modified from \citet{Salyk13} (see text).  Since
large systematic error ($\gtrsim 1$ dex) exists when comparing accretion rates
from different indicators, we have not connected the open symbols to the line
connecting the \Brg derived points.\label{mdotsFig}}
\end{center} 
\end{figure}

In Figure \ref{mdotsFig} we plot the accretion rate versus time for each source
in the binary. The additional \Brg derived accretion rates for UY~Aur~A were
calculated using the EWs from \citet{Herbst95} and \citet{Fischer11} and the
continuum flux level observed by \citet{Close98}.  The flux level for UY~Aur~B
is always deduced by comparing to UY~Aur~A.  Since \citet{Herbst95} and
\citet{Fischer11} do not report errors for their EW measurements we assign them
our measured value of 0.3 \AA. This is probably conservative since most of our
uncertainty comes from defringing. We use 0.05 magnitudes of uncertainty in the
continuum flux level, the value reported in \citet{Close98}.  Some additional
uncertainty in the flux level is accrued by using flux calibration data not
collected in parallel with the spectroscopic observations.  For example, young
binary stars as a class can vary by $\gtrsim0.4$ magnitudes in the NIR
\citep[e.g][]{Skrutskie96,Hioki07}. However, since the flux ratio between A and
B is measured, any untracked variability in the flux of UY~Aur~A will affect
both components of the binary similarly, preserving the relative trends.  In
addition to \Brg derived accretion rates, we also include measurements using
other techniques in Figure \ref{mdotsFig}.  With triangle and square symbols we
show accretion rates derived using both optical veiling measurements
\citep{Hartigan03}, and Pf$\beta$ measurements \citep{Salyk13}, respectively.

The published Pf$\beta$-derived accretion rates \citep{Salyk13} appear to
be at odds with the long-term trends in accretion rate inferred from \Brg and
optical veiling measurements.  A re-analysis of the NIRSPEC data suggests that
the components were misidentified, and we correct that here.  However, we
caution that the NIRSPEC slit was misaligned with the binary PA during these
observations, and this adds systematic uncertainty to the relative accretion
luminosities derived from these data.

\begin{figure}[h!]
\begin{center}
\epsscale{0.8}
\plotone{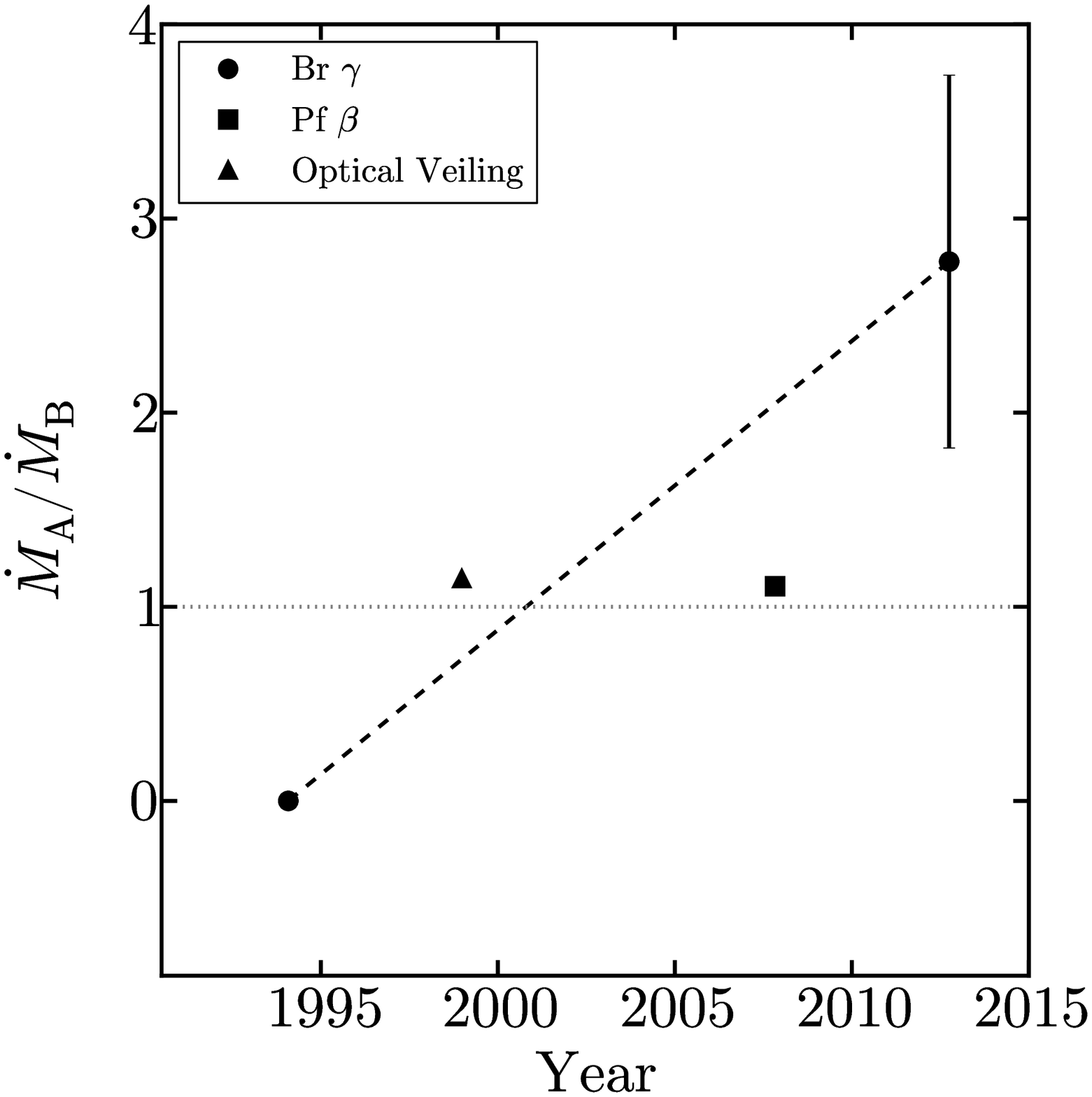}
\caption{The relative accretion rate,
$\dot{\mathrm{M}}_{\mathrm{A}}$/$\dot{\mathrm{M}}_{\mathrm{B}}$, for each epoch
where the rate was measured for both sources. Accretion rate indicators are not
corrected for extinction, which can be variable. The accretion rate ratios
derived using \Brg luminosity have been connected by a dashed line. A dotted
line indicates a ratio of unity. UY~Aur~A now appears to be accreting at
a faster rate than UY~Aur~B.\label{mdotsRatioFig}} 
\end{center}
\end{figure}

In Figure \ref{mdotsRatioFig}, we show the accretion rate ratio versus time for
all epochs when the accretion rate was measured for both sources. Figure
\ref{mdotsRatioFig} shows that accretion was directed primarily toward the
secondary component in the initial epoch, but transitioned to the primary over
time. The timescale of the observed variability is much shorter than the
$\sim2000$ year dynamical timescale of the binary. Rather, the 5-10 year
timescale corresponds to the orbital timescale at 2.5--5 AU from the
$\sim0.5\mathrm{M}_{\odot}$ stars.

\begin{deluxetable*}{rlllllllll}
\tabletypesize{\scriptsize}
\tablecolumns{10}
\tablewidth{0pt}
\tablecaption{Equivalent Widths and Line Luminosities}
\tablehead{
            \colhead{UY~Aur} &
            \colhead{$\mathrm{m}_{\mathrm{K}}$\tna}&
            \colhead{EW(\Brg)}&
            \colhead{$\mathrm{L}_{\mathrm{Br}\gamma}$\tnb}&
            \colhead{$\mathrm{L}_{\mathrm{acc}}$\tnc}&
            \colhead{$\mathrm{\dot{M}}$\tnd}&
            \colhead{EW(S(1))}&
            \colhead{$\mathrm{L}_{S(1)}$}&
            \colhead{EW(S(0))}&
            \colhead{$\mathrm{L}_{S(0)}$}
            \\
            \colhead{ } &
            \colhead{ } &
            \colhead{[\r{A}]}&
            \colhead{$[\mathrm{L}_{\odot}]$}&
            \colhead{$[\mathrm{L}_{\odot}]$}&
            \colhead{$[\mathrm{M_{\odot}}\mathrm{yr}^{-1}]$}&
            \colhead{[\r{A}]}&
            \colhead{$[\mathrm{L}_{\odot}]$}&
            \colhead{[\r{A}]}&
            \colhead{$[\mathrm{L}_{\odot}]$}
           }
\startdata
A & $7.42$ & $-2.7(0.3)$  & $7.3(0.8)\mathrm{E}$$-5$ & $1.6(0.2)\mathrm{E}$$-1$ & $3.5(0.5)\mathrm{E}$$-8$ & $-0.20(0.02)$ & $5.4(0.5)\mathrm{E}$$-6$ & $-0.06(0.01)$   & $1.6(0.4)\mathrm{E}$$-6$ \\
B & $7.67$ & $-1.1(0.3)$  & $2.3(0.5)\mathrm{E}$$-5$ & $0.4(0.1)\mathrm{E}$$-1$ & $1.2(0.4)\mathrm{E}$$-8$ & $-0.08(0.01)$ & $1.7(0.2)\mathrm{E}$$-6$ & $-0.025(0.012)$ & $0.5(0.2)\mathrm{E}$$-7$ 
\enddata
\tablenotetext{a}{The K-band magnitude for UY~Aur A is from \citet{Close98}.
For UY~Aur B, we determined the magnitude difference using our own measured
flux ratio.}
\tablenotetext{b}{\Brg luminosities are not corrected for extinction, following
the approach of previous studies. We discuss below the potential effects of
strong extinction on the \Brg line of UY Aur B.}
\tablenotetext{c}{Accretion luminosity is derived using the correlation
reported by \citet{Muzerolle98}}
\tablenotetext{d}{Mass accretion rates were derived using the stellar radii and
masses reported in \citet{Hartigan03}}
\tablecomments{The uncertainty in each reported value is presented in
parentheses. Luminosities were derived assuming a distance of 140 pc.
\label{EWTable}}
\end{deluxetable*}

\subsection{\HH Lines}
\begin{figure*}
\begin{center}
\epsscale{0.8}
\plotone{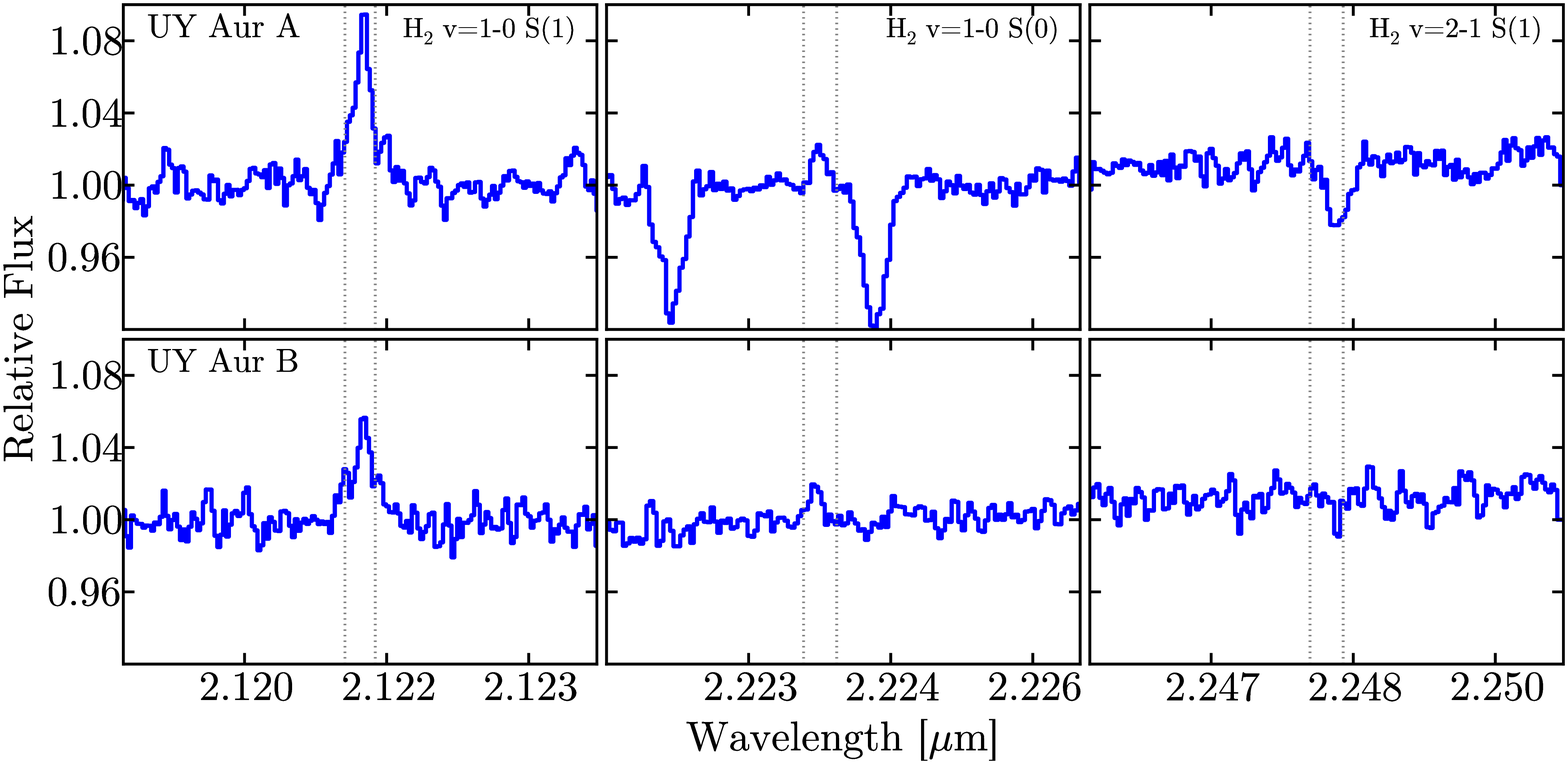}
\caption{Molecular Hydrogen lines in the spectra of both UY~Aur~A (top), and
UY~Aur~B (bottom). The v=1-0 S(1) and S(0) lines are detected in both sources,
while the v=2-1 S(1) line is detected in neither. However, a photospheric
absorption line in the spectrum of UY~Aur~A appears in the spectrum near where
we would expect the v=2-1 S(1) line. We note that each order was independently
normalized to make this figure. \label{H2Fig}}
\end{center}
\end{figure*}

In Figure \ref{H2Fig}, we show portions of the stellar spectra near the
molecular Hydrogen v=1-0 S(0) and S(1) lines and the v=2-1 S(1) line.  We
detect the v=1-0 S(0) and S(1) lines in both sources.  A photospheric
absorption line in the spectrum of \UYA complicates our ability to place
a strong constraint on the v=2-1 S(1) line. We do not detect the line in the
spectrum of UY~Aur~B. 

We have calculated the v=1-0 S(1) to S(0) ratio for each source. The ratios and
associated uncertainty intervals are plotted in Figure \ref{H2Fig}. We have
also converted these ratios to excitation temperatures using the transition
probabilities of \citet{Wolniewicz98}. The implied temperatures, $\sim1025$
K and $\sim750$ K for \UYA and B, respectively, are also plotted in Figure
\ref{H2Fig}. At these temperatures the v=2-1 S(1) line should be very weak,
$\lesssim 1\%$ the intensity of the v=1-0 S(1) line. This is compatible with
the absence of strong v=2-1 S(1) emission in our spectra. Since models of
fluorescent diffuse gas predict that the strength of the v=2-1 S(1) line should
be similar to, even greater than, the v=1-0 S(0) line \citep[e.g.][]{Black87},
we can rule out a large contribution to the \HH line emission from fluorescent
diffuse gas.

There are several ways to produce a thermalized spectrum of \HH lines.
Mechanisms include shock heating and UV/X-ray irradiation of dense gas. Shock
heating, in an inflow or outflow, is expected to produce relatively broad lines
($\gtrsim 100$ km/s) offset from the continuum \citep[e.g.][]{Beck08,Beck12}.

We observe lines that are unresolved (widths $<20$ km/s) and not offset in
velocity from the stellar photospheres. We detect no astrometric displacement
along the slit (PA=40$^{\circ}$). Our measured astrometric precision along the
slit is $\sim0.5$ AU. Given the observed v=1-0 S(1) line-to-continuum ratio for
UY~Aur~A, $\sim0.1$, we constrain the bulk of the observed \HH flux to be
centered within 15 AU of the continuum at the 3-$\sigma$ level. Furthermore, we
see no significant increase in the line-to-continuum ratio of the \HH lines
when we extract spectra from the margins of our slit, $0.3''-0.5''$ ($42-70$ AU)
from the continuum centroid. In these off-star extractions we can
constrain any extended emission in the $0.3''-0.5''$ region, along a PA of
40$^{\circ}$, to be less than $\sim1\times10^{-13}$ ergs cm$^{-2}$ s$^{-1}$
arcsec$^{-2}$. Since we do not detect any spatial offsets or extension in
the \HH lines, it is difficult for us to directly implicate shocks with these
data.

Assessing whether FUV or X-ray irradiation could be responsible for excititing
the \HH lines is difficult because no well characterized FUV spectrum exists
for UY~Aur. \citet{Nomura07} used the spectrum of TW~Hya to model the \HH
ro-vibrational spectrum from a proto-planetary disk around a 0.5 M$_{\odot}$
star. For the case with the smallest grains, they predict thermal line ratios
and a v=1-0 S(1) line strength 2.5 times as strong as what we report here for
UY~Aur~A. Since \UY has a weaker X-ray luminosity than TW~Hya \citep[by
a factor of $\sim$ 5,][]{Guedel2010}, it is not surprising that \UYA has
a somewhat weaker line strength than predicted by the \citet{Nomura07} model.
High density UV fluorescence combined with X-rays seems capable of generating
much of the observed compact \HH emission.

\begin{figure}[h!]
\begin{center}
\epsscale{0.8}
\plotone{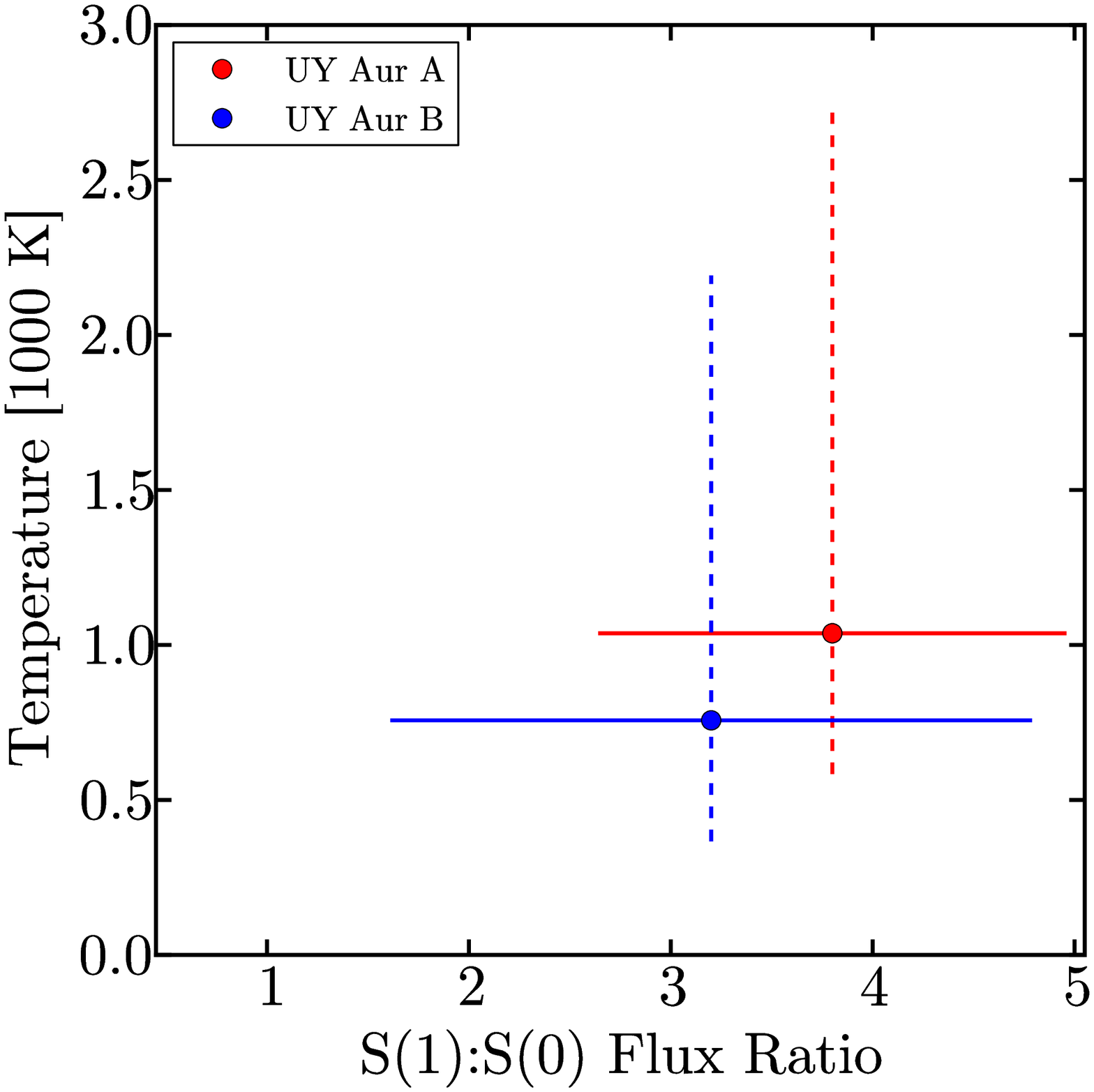}
\caption{The molecular gas temperature deduced from the \HH line ratios for
each source, using an LTE model. The symbol for each source indicates the
measured ratio and deduced temperature. The horizontal error bars indicate the
range of ratios allowed given our uncertainty in the EW ratio of the lines. The
vertical errorbars indicate the corresponding range of
temperatures.\label{tempFig}} 
\end{center} 
\end{figure}

\subsection{Veiling}
Veiling is the ratio of excess flux ($E_{\lambda}$) over photospheric flux
($P_{\lambda}$) at a given wavelength: 
\begin{equation}
r=\frac{E_{\lambda}}{P_{\lambda}}.  
\end{equation} 

\begin{figure*}[h!] \begin{center}
\epsscale{0.8}
\plotone{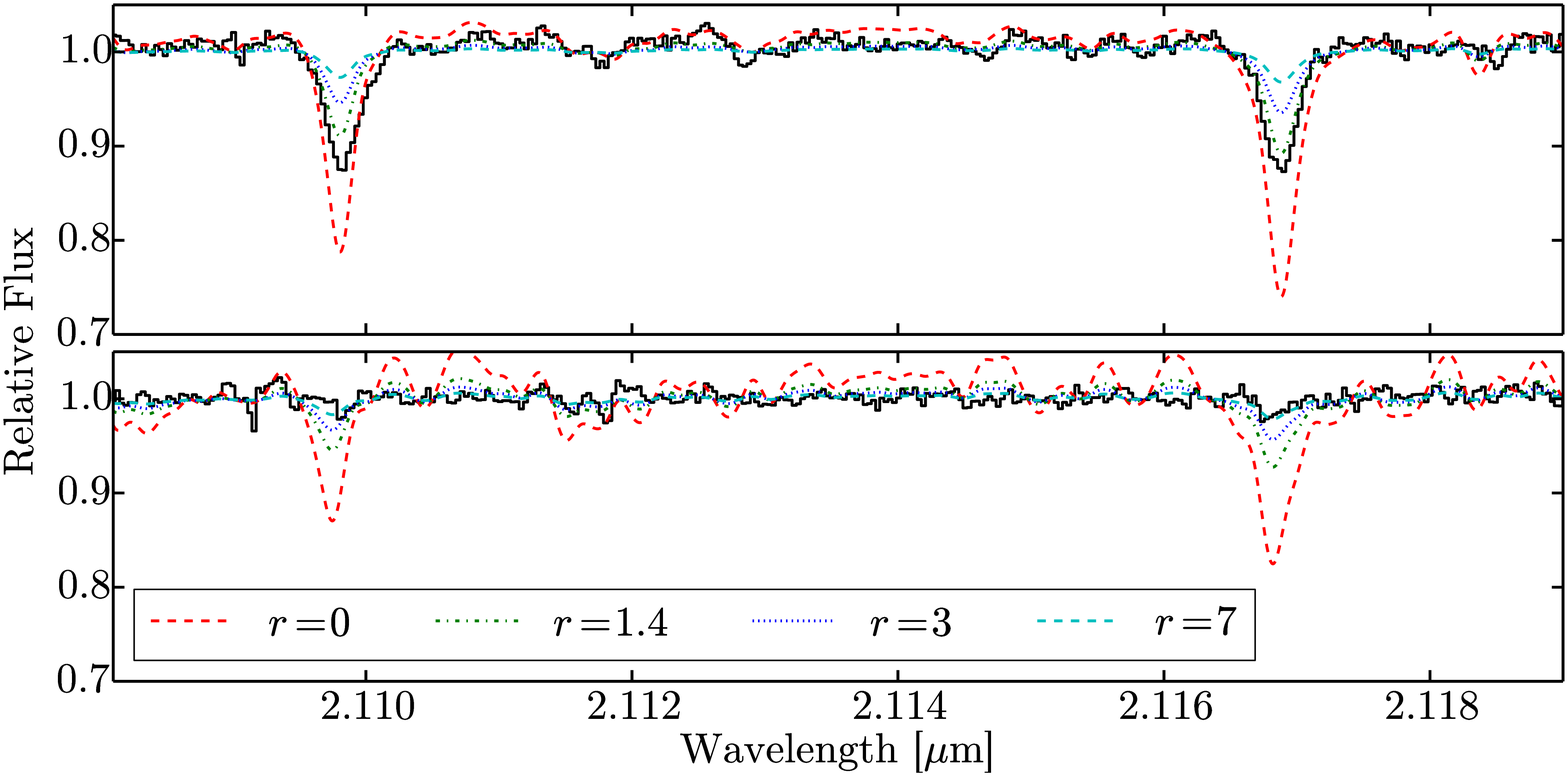}

\caption{ Top: A region of the observed spectrum of UY~Aur A showing two
Aluminum photospheric absorption lines (black histogram). We overplot a PHOENIX
model photosphere (T$_{\mathrm{eff}}$=3900K, logg=4.5) veiled by $r=0$ (dashed,
red), 1.4 (dot dashed, green), 3 (dotted, blue), and 7 (dashed, cyan). The
$r=1.4$ value was reported by \citet{Fischer11} for UY~Aur~A, and closely
matches our observations.  Bottom: UY~Aur B observed spectrum and veiled
synthetic spectra (T$_{\mathrm{eff}}$=3900K, logg=4.5). Photospheric lines in
the spectrum are nearly absent, a significant change from the observations
reported by \citet{Herbst95}.\label{VeilFig}} 

\end{center} 
\end{figure*}

Since we did not observe photospheric calibrators for UY~Aur~A and B, making
a quantitative measurement of the veiling toward each source is difficult.
Without a photospheric template for each source we cannot characterize internal
instrumental scattered light, which can artificially veil our spectra (see
\citet{Eisner10}, for an example of instrumental veiling with ARIES).  However,
we can make a qualitative statement about the relative veiling between each
source. In Figure \ref{VeilFig} we show a portion of the spectra of both
UY~Aur~A (top panel) and UY~Aur~B (bottom panel) in the vicinity of two strong
Aluminum photospheric absorption lines. We also plot synthetic PHOENIX
photospheric spectra \citep{Husser13}. The model spectra represent stars with
logg=4.5 and T$_{\mathrm{eff}}$=3900~K and T$_{\mathrm{eff}}$=3500~K for
UY~Aur~A and B, respectively.  We veiled the synthetic spectra by $r=0$, 1.4,
3, and 7.  The $r=1.4$ value is that reported by \citet{Fischer11} for UY~Aur
A.  As shown in the figure, $r=1.4$ matches our observation for UY~Aur~A well,
suggesting that instrumental veiling is weak or absent.  Our spectrum of
UY~Aur~B shows significantly weaker absorption lines than the $\sim10\%$-depth
lines observed in the R=250 spectrum reported by \citet{Herbst95}. Thus we
report that a large change in the K-band veiling of UY~Aur~B has taken place
since 1994.

\section{Discussion}\label{Disc} 

In the previous section we presented the second ever K-band spectrum of
UY~Aur~B. We observe a $\sim0.9$ magnitude increase in the flux of this source,
together with increased veiling and decreased emission line luminosity. An
increased circumstellar disk flux could explain both the increased K-band flux
and the stronger veiling. Such an increase could be due to an episodic increase
in the mass accretion rate through the disk. However, we must
understand why the observed \Brg luminosity, a tracer of accretion onto the
star, has decreased even while accretion through the disk has increased.

One explanation appeals to a nearly edge-on viewing geometry to explain the
increased luminosity of the disk and the decreased luminosity of the \Brg line.
Several observations of the system hint at an edge-on orientation, including
the unusually discrepant visual extinctions toward UY~Aur~A and
B \citep{Hartigan03}, the extincted appearance of the 10-micron silicate
feature \citep{Skemer2010}, and the large amplitude optical and NIR variability
\citep[e.g. compare][]{Joy44,Herbst95,Close98,Brandeker03,Hioki07}. In this
scenario, when the scale height of the disk responds to the increased accretion
rate and temperature ---becoming larger--- our view of the star becomes more
obscured by the inner disk rim. Thus, the intrinsic \Brg luminosity can
increase but be observed to decrease.

We estimate the level of increased K-band extinction necessary to produce
the observed change in the \Brg luminosity given an assumed factor for the
increased disk luminosity. Measuring the increase in disk luminosity requires
a previous quantitative measurement of the K-band veiling for UY~Aur~B, which
does not exist. We can make a ballpark estimate assuming that the previous
veiling was $r_{\mathrm{K}}\lesssim1$, a common value for other classical
T Tauri stars \citep{Fischer11}, and qualitatively consistent with the previous
K-band spectrum \citep{Herbst95}. For example, if the 0.86 magnitudes increase
we observe in the K-band flux is driven by a factor of 3 increase in the
luminosity of the disk, and we assume a similar increase in the intrinsic
accretion luminosity of material making it onto the star, then we expect
a factor of 2.4 increase in the intrinsic \Brg luminosity \footnote{Using the
correlation between accretion luminosity and \Brg luminosity in
\citet{Muzerolle98}}. However, the observed \Brg line strength has decreased
by a factor of 0.54. This implies that only 23\% of the \Brg luminosity is
passing through the puffed up disk rim, an increase in A$_{\mathrm{K}}$ of 1.6.

We convert such an increase in A$_{\mathrm{K}}$ to a change in column density
along the observed line of sight using a value for the K-band dust opacity.  We
use $\kappa_{\mathrm{K}}=10$ cm$^{2}$g$^{-1}$, compatible with models of
T Tauri disk dust grains \citep{Miyake93}, to derive the required column
density increase: 0.15 g/cm$^{2}$. The fractional increase implied by such
a change depends on the initial column density along our line of sight. Small
fractional changes are required along lines of sight initially transecting
optically thick portions of the inner rim while larger fractional changes are
required along sight lines through more tenuous material. For example, if our
sight line radially through the rim initially transected where $\tau=1$, then
a 3\% change in the scale height of the disk is necessary.  A 10\% change in
the scale height is required if our initial line of sight was through the rim
at a height where $\tau=0.2$. Using L$\propto$T$^{4}$, and assuming
a hydrostatic disk, h$\propto$c$_{s}$$\propto$T$^{0.5}$, we expect a 14\%
increase in the scale height given our assumed increase in the luminosity of
the disk. 

Alternatively, if the observed decrease in the \Brg flux and the increased disk
luminosity implies an intrinsically lower accretion rate onto the central star,
material flowing through the disk must accumulate at some radius or be ejected
in an outflow before making its way onto the star.  In a viscous accretion
disk, mass is expected to pile up in regions with relatively low viscosity.
\citet{Gammie96} described an accretion flow where the mid-plane of the disk at
intermediate radii has very low viscosity because the gas, which is neutral
there, is decoupled from the magnetic field. Mass is predicted to accumulate in
these ``deadzones". \citet{Gammie96} pointed out that such an accumulation of
mass would lead to instability, resulting either in the fragmentation of bound
objects (e.g.  planets) or in sufficient heating of the deadzone to ionize
enough of the gas to couple the material there to the magnetic field, causing
a sudden and dramatic increase in the accretion rate.  If matter is indeed
piling up in a deadzone, then we would expect \UYB to undergo outbursts in
the future.

One clue that can help us understand the intrinsic accretion properties of
UY~Aur~B is the observation of extended outflows. Mass-loss rates from young
stars are correlated with mass accretion rates \citep[e.g.][]{Hartigan95}, and
outflows can be used as a tracer of accretion. Recently, \citet{Pyo14} reported
the observation of extended outflows surrounding UY~Aur.  These extended
outflows should not be extincted by circumstellar disk material at small radii,
and may give a clearer perspective on the intrinsic accretion rate of UY~Aur~B.
Thus, time monitoring of outflow rates from \UYB will help to constrain the
physics of the variability in this source.

Additionally, both of the \HH excitation scenarios discussed above (shock or
UV/X-ray excitation) should produce \HH lines with intensity that correlates
with the accretion rate. For shocks, a larger accretion rate should result in
a larger outflow rate \citep[e.g.][]{Hartigan95}, strengthening the \HH signal.
An increased accretion rate should also produce a stronger UV and X-ray
accretion-shock spectrum and strengthen the fluorescent excitation. Thus our
observed decrease in \HH line strength compared to the observations by
\citet{Herbst95} may indicate a decrease in the intrinsic accretion rate. 

If the highly inclined geometry explanation for the behavior of \UYB is
correct, we predict that veiling, total flux, and outflow tracers of accretion
should be correlated, but that shock-tracers of accretion (such as H I lines
like \Brg) should be anticorrelated.  More frequent monitoring of this system
will provide better knowledge of the nature and timescale of the variability
and will be important in disentangling the physical state of UY~Aur~B.

Since the circumstellar disk around \UYA is not edge on \citep{Akeson14}, we
predict that all tracers of accretion, veiling, and total flux should be
correlated. Three epochs of published spatially separated K-band spectroscopy
exist for UY~Aur~A. The accretion rate seems to be increasing
$\sim2\times10^{-9}$ M$_{\odot}$yr$^{-1}$ per year.  \UYA now has the larger
observed \Brg luminosity.  In addition to the indicators included in our Figure
\ref{mdotsFig}, further evidence that \UYA has been trending toward a more
active state can be seen in the comparison of the He I line shapes of
\citet{Edwards06} and \citet{Pyo14}. In 2014, \citet{Pyo14} observed deeper red and
blue shifted absorption features than \citet{Edwards06}. This suggests more
vigorous infall and outflow from the source. 

Additional evidence that UY~Aur~A is now in an intrinsically more active state
than UY~Aur~B comes from the relative outflow rates. \citet{Pyo14} showed that
UY~Aur~A was driving an outflow at a larger rate than UY~Aur~B in 2007,
consistent with our Figure \ref{mdotsFig}.  However, if we assume the outflow
is launched at the velocity observed by \citet{Pyo14} in the [FeII] lines,
$\sim150$ km/s, then we expect $\sim2$ year delay between variations in
accretion at the stellar surface and variations in the observed outflow. 

If UY~Aur~B is in fact accreting less than UY~Aur~A then we must understand why
the more vigorous accretor in UY~Aur seems to alternate between the primary and
the secondary. Theory predicts that circumbinary disks should preferentially
either the primary, as shown by \citet{Ochi05}, or the secondary as predicted
by\citet{Bate97}. One solution, which is consistent with the data, is that the
accretion onto both sources, as traced by Br$\gamma$, is episodic.  In this way
the average accretion rate onto the primary or secondary could dominate over
time. This scenario has significant implications for planet formation because
intense bursts of accretion can alter the chemical make-up and particle
distribution in the disk by dissociating molecules and evaporating dust. 

Another option is that the UY~Aur binary system is substantially different from
modeled systems.  For example, \citet{Bate97} and \citet{Ochi05} modeled
systems with the implicit assumption that the binary seeds result from a disk
fragmentation process. The possibly mis-aligned inclinations of the
circumstellar disks in \UY is a clue that \UY may not have formed via disk
fragmentation. This more complicated scenario is not as easily compared to
clear-cut theoretical predictions.

\section{Conclusions} 

We have presented R=15,000 K-band spectroscopy of each star in the UY~Aur
binary system. Our spectrum of UY~Aur~B is only the second ever published and
enables time domain studies at wavelengths probing both
circumstellar material and accretion. We highlight significant changes since
1994.  Specifically, we see that UY~Aur~A presents increased \Brg luminosity
and now shows \HH emission while UY~Aur~B shows a decrease in the strength of
Br$\gamma$ and \HH emission. In contrast to the state of the system in 1994, UY~Aur~A
now has the stronger observed \Brg line.  Additionally, UY~Aur~B now appears
significantly brighter and more heavily veiled than it has in the past.

We suggested two scenarios to explain the observed changes in the state of
UY~Aur~B. Both cases include increased viscous accretion through the
circumstellar disk to account for the increased luminosity and stronger veiling
observed. In the first scenario a similar increase in the \textit{intrinsic}
\Brg line strength is attenuated by increased extinction due to a puffed-up
edge-on disk. We showed that the change in the scale height of the disk that is
necessary to produce the reduced \Brg line strength is compatible with the
amount of change expected for a disk undergoing an episode of increased viscous
accretion.  Alternatively, the observed reduction in the \Brg line strength,
and the accretion onto the star which it traces, is intrinsic. In this case,
mass flowing through the disk must pile up or be ejected before it reaches the
star.  We predict, if the edge-on disk scenario is correct, that future higher
cadence time-monitoring of this system will reveal an anti-correlation in the
variability of shock-tracers of accretion (such as H I lines) and the observed
disk luminosity and veiling.

\section{Acknowledgments}
This work was supported by NASA Origins grant NNXX11AK57G.

\clearpage

\end{document}